\documentclass[]{article}
\usepackage{amssymb,amsfonts,amsthm,bm,graphicx,amsmath,float,color,longtable}
\restylefloat{figure}
\usepackage[ansinew]{inputenc}
\usepackage{authblk}

\usepackage{crop}%
\usepackage{amsmath,amssymb,amsfonts,upref,endnotes}
\usepackage{amsthm}
\usepackage{marginnote}
\usepackage{soul}
\RequirePackage{graphicx}
\usepackage[figuresright]{rotating}
\usepackage{color}

\usepackage{fancyhdr}
\usepackage{lastpage}
\usepackage[scaled]{helvet}
\usepackage[T1]{fontenc}

\definecolor{jobcolor}{cmyk}{1,.81,.09,.01}
\usepackage{xcolor}
\usepackage{sectsty}
\sectionfont{\color{jobcolor}}  

\usepackage{helvet,times}
\usepackage{bm,textcomp}
\usepackage{amsmath,amsthm,amssymb,amsbsy,epsfig,fancyhdr,calc,ifthen,float,psfrag,mathrsfs,graphics,color,fullpage}
\restylefloat{figure}
\usepackage{subcaption} 
\usepackage[ansinew]{inputenc}

\usepackage{empheq}

\begin{document}

\title{Load-dependent adaptation near zero load in the bacterial flagellar motor}

\author[1,2]{Jasmine A. Nirody}
\author[3]{Ashley L. Nord}
\author[1]{Richard M. Berry}
\affil[1]{Clarendon Laboratory, Department of Physics, University of Oxford, Oxford, UK}
\affil[2]{Center for Studies in Physics and Biology, The Rockefeller University, New York, USA}
\affil[3]{Centre de Biochimie Structurale,  INSERM, CNRS, Universit\'{e} de Montpellier, Montpellier, France}
\date{}

\maketitle

\begin{abstract} 
{The bacterial flagellar motor is an ion-powered transmembrane protein complex which drives swimming in many bacterial species. The motor consists of a cytoplasmic `rotor' ring and a number of `stator' units, which are bound to the cell wall of the bacterium. Recently, it has been shown that the number of functional torque-generating stator units in the motor depends on the external load, and suggested that mechanosensing in the flagellar motor is driven via a `catch bond' mechanism in the motor's stator units. We present a method that allows us to measure -- on a single motor -- stator unit dynamics across a large range of external loads, including near the zero-torque limit. By attaching superparamagnetic beads to the flagellar hook, we can control the motor's speed via a rotating magnetic field. We manipulate the motor to four different speed levels in two different ion-motive force (IMF) conditions. This framework allows for a deeper exploration into the mechanism behind load-dependent remodelling by separating out motor properties, such as rotation speed and energy availability in the form of IMF, that affect the motor torque.}

{
Correspondence: jnirody@rockefeller.edu\\
}
\end{abstract}

The bacterial flagellar motor (BFM) is an ion-driven nanomachine that drives swimming in a variety of bacterial species. The BFM couples the flow of cations (protons, in \textit{Escherichia coli}) across the bacterial membrane to induce rotation in the flagellum, spinning the filament like a propeller to move the bacterium forward \cite{Berg2003,nirody2017biophysicist}. The flagellar motor generates torque through interactions between the motor's stator and rotor; specifically torque is generated via an interaction between a stator unit (in \textit{E. coli}, comprising the proteins MotA and MotB) and FliG protein `spokes' that line the rotor's cytoplasmic C-ring \cite{mandadapu2015mechanics,chang2019structural} (Figure 1a). The BFM's stator can be composed of between 1 and at least 11 independent units \cite{block1984successive,samuel1996torque,leake2006stoichiometry,Reid2006}.

The flagellar motor plays a crucial role in bacterial pathogenicity and in several processes like chemotaxis and biofilm formation. Though understanding its function has been a long-standing problem in biophysics, its size and localisation to the membrane have 
made this challenging. The development of the tethered cell assay definitively confirmed that the BFM was a rotary motor \cite{silverman1974flagellar}. In this assay, cells were tethered to a surface by a single filament, and the motor rotation was characterised via rotation of the tethered cell. This assay is ideal for direct observation of rotation by eye, as the load of the cell body is such that the motor can rotate at only a few Hertz. However, experiments using tethered cells limited measurements of motor rotation to high loads. Later, the bead assay, in which beads of varying sizes, and accordingly varying drags, are attached to flagellar stubs of adhered cell bodies, allowed motor rotation to be studied across a wider range of loads \cite{ryu2000torque}.

Beyond the torque-speed relationship in `fully assembled' motors, the discovery that motor components are constantly being turned over opened up a new avenue of scientific questions. In particular, stator units have been shown to continuously exchange with a large pool of membrane-bound `spares' \cite{leake2006stoichiometry,sowa2008bacterial}. Further evidence that motors at high loads are able to maintain a much higher number of engaged stator units than motors at low load may suggest that bacterial mechanosensing, utilised in the transition from liquid to surface living, may arise from the flagellar motor \cite{Lele2013,Tipping2013} .

This discovery both pointed to an exciting new line of research to be explored in the future, and brought into question the results of several experiments from the past. For one, it meant that all previously measured torque-speed relationships were likely made on motors with varying numbers of stator units. This has been foundational to the development of theoretical models developed since \cite{mandadapu2015mechanics,nirody2016limiting}. As the motor's structure and function was shown to vary with external conditions, tools to manipulate the load on single motors proved vital to properly understand remodelling in the BFM. This has been achieved through monitoring the process of cell tethering \cite{Lele2013}, which dramatically increases the load on the motor, and electrorotation \cite{sato2019evaluation,wadhwa2019torque}, which uses a rotating electric field to apply external torque on the motor.

In a recent paper, the molecular mechanism behind stator remodelling was investigated using an external magnetic field and magnetic microbeads of different sizes bound to the hook of individual \textit{E. coli} motors \cite{nord2017catch}. Load-dependent remodelling in the BFM can ostensibly manifest in several ways: by stator units engaging at a higher rate ($k_\text{on}$) or disengaging at a lower rate ($k_\text{off}$) as a function of the load, or by some combination of the two. Nord et al. showed evidence that $k_\text{off}$ (and thus unit lifetime) is load-dependent, suggesting that the dynamics behind stator assembly are consistent with a catch bond \cite{nord2017catch}. This somewhat counterintuitive bond is one that is strengthened, rather than weakened, as more force is applied to it --- stator units stay engaged onto the motor longer at higher loads.

Nord et al. used permanent magnets to stall motors with magnetic microbeads attached to their hooks. During the period when the motor is stalled, additional stator units are recruited to the motor. Upon release, units disengage and the motor relaxes to its steady state structure, determined by the load imposed on the motor by the drag of the bead \cite{nord2017catch}. This method allowed for exploration of stator dynamics at the full range of loads accessible by the bead assay, but not those near vanishing load. In the following, instead of permanent magnets, we use a rotating electromagnetic field, which allows us to rotate the motor across the entire range of speeds achievable by the BFM. Our experiments were contemporary with, and are complementary to, a similar investigation using electrorotation \cite{wadhwa2019torque}. We discuss electrorotation as an alternate method to manipulate motor torque on the flagellar motor in the Discussion.

Motor torque is a complex function of several factors, including motor speed and ion-motive force (IMF) \cite{fukuoka2009sodium,lo2013mechanism,Tipping2013}. Here, we probe the dynamics behind load-dependent motor remodelling, focusing on the low-load regime, where motor torque varies most dramatically with speed (Figure 1). We also investigate how these dynamics depend on ion-motive force, by measuring stator remodelling at the same speeds as IMF is varied, and provide support that mechanosensing in the BFM shows behaviour akin to that of a catch bond across a wide range of motor operating conditions \cite{nord2017catch}.

\section*{Manipulating motor speed using rotating magnetic field}

Using a rotating magnetic field, we control the speed of single motors by exerting torque on a superparamagnetic bead attached to a flagellar hook (Figure 2a). The torque experienced by the bead from a magnetic field $\overrightarrow{B}$ is given by
\begin{equation}
\overrightarrow{\tau_B} = \overrightarrow{m} \times \overrightarrow{B}.
\end{equation}
The applied torque $\overrightarrow{\tau_B}$ on the bead will be nonzero only if the magnetic dipole moment $\overrightarrow{m}$ is not aligned with $\overrightarrow{B}$. These beads comprise randomly oriented anisotropic magnetic nanoparticles (NP), and $\overrightarrow{m}$, the vector sum of the dipole moments of the individual NPs, is in general not parallel with $\overrightarrow{B}$ allowing for torque generation \cite{van2015biological}. When the magnetic field $\overrightarrow{B}$ is rotated at speed $\omega_B$, the superparamagnetic bead will follow at the same speed so long as the maximum applied torque $\overrightarrow{\tau_B}$ is sufficient to satisfy the balance between $\overrightarrow{\tau_B}$, the motor torque, and the drag of the bead (Figure 2b).

To generate the rotating field $\overrightarrow{B}$, a three-pole electromagnet was positioned over a sample on the stage of an inverted microscope; this design is a modification of the six-pole magnet system constructed in \cite{gosse2002magnetic}. The core of the magnet was constructed of a highly magnetic-permeable ferrite core. Three removable soft-iron poles were attached to slots within an iron block, which was attached to the core; this configuration allowed us to vary the spacing of the gap between the pole pieces (and thus the magnitude of the magnetic field at the centre of the magnet) while strictly maintaining a three-fold symmetry, which allows rotation of the direction of the magnetic field and therefore of the magnetic particle \cite{gosse2002magnetic}.

An enamelled 20 British Standard Wire Gauge (SWG) copper wire (diameter$\sim$1 mm) was used to wind three separate coils, each composed of 108 turns, around the electromagnet. A polyimide film (Kapton, DuPont) was used between wire layers and at the ends of each coil for electrical and thermal insulation. Each coil was driven by a voltage-to-current amplifier. Voltage-output modules (N19263, National Instruments) were used as a signal generator to modulate inputs into the amplifiers; inputs to the three coils were sinusoids of equal amplitudes and 120$^{\circ}$ out of phase. Further details on magnet construction and calibration can be found in \cite{gosse2002magnetic} and \cite{lim2015application}.

In our assay we use a non-switching strain of \textit{E. coli} (MTB32 + $\Delta$CheY) with an endogenously biotinylated hook \cite{brown2012flagellar,nord2017catch}. Streptavidin-coated superparamagnetic beads (MyOne DynaBeads T1, diameter 1$\mu$m) were attached to the biotinylated hooks after cells were immobilised onto tunnel slides using poly-L-lysine. Bead position was determined at 10k Hz using back-focal plane interferometry and a quadrant photo diode (QPD; see, e.g., \cite{sowa2010simple,nord2017speed}). Bead $(x,y)$ positions from the QPD were converted to speed vs time traces using a Fast Fourier Transform (FFT) with a sliding zero-padded window of length 0.2s. A median filter with window 0.5s was then applied to the speed traces. All code was written in Python and is available at http://users.ox.ac.uk/$\sim$phys1213.

Motors were allowed to rotate for at least 60s before current was passed through the magnets and torque was applied.  Stator unit number was determined from jumps in the first 60s of speed traces as in \cite{nord2017catch}; the single-unit speed (3.7 $\pm$ 1.0 Hz in the normal condition and 1.9 $\pm$ 0.9 Hz in the butanol condition) determined from these discontinuities was used to quantify stator unit stoichiometry for the remainder of the trace. This assumption, that the motor speed increases linearly with stator number, and thus that stator number can be quantified by the speed of a single step, has been generally accepted at high loads \cite{sowa2008bacterial,nirody2016limiting,nord2017speed}, where our stoichiometry estimates are made. An example trace is shown in Figure 3a.

Motors were then rotated at fixed speed by flowing a current through the three-pole electromagnet, as shown in Figure 1. Each `magnet sequence' consists of ten 5s intervals of forced rotation by the electromagnet at a fixed speed, separated by 0.2s intervals with the magnets off, during which the number of stators was determined (Figure 3a,b). Rotation speeds were chosen as ($nx +1$) Hz, ${n=1,2,4,6\}}$, where $x$ is the acquisition rate of the camera (50 Hz) used to observe the brightfield video. This allowed easy visual determination of whether the bead was rotating at the same speed as the magnetic field: such rotation was aliased to 1 Hz by the camera sampling (see supplementary video).

Currents of amplitude 1.5A were used for all experiments; this strength allowed us to reach zero-torque speeds at a sufficient yield using commercially available beads while minimising any force on the motor \cite{gosse2002magnetic}. To further assure that the beads' moments were aligned with the magnetic field, beads were flowed into the tunnel slide and left to settle in the presence of a magnetic field of low strength, generated using a pair of Helmholtz coils. This ensured that free beads were reasonably well-aligned before attaching to a flagellar hook. All experiments were performed at 25$^{\circ}$C in motility buffer (10mM potassium phosphate, 0.1mM EDTA, pH 7.0) with or without the addition of 0.5\% butanol by volume. 

Within each magnet sequence, periods of applied torque are called `magnet on' periods, and intervals in between (during which the motor is allowed to rotate freely) are called `magnet off' periods. The motor speed was counted as the average speed in the 0.2s directly after the current to the magnets was shut off (see Figure 3a). Recordings from `magnet off' intervals were discarded and treated as missing data if the motor did not maintain a speed within 5 Hz of the speed of the rotating magnetic field for at least 4 consecutive seconds in the preceding `magnet on' interval (i.e., segments in which the bead did not sufficiently follow the magnetic field's rotation). Motors were allowed to return to within 5 Hz of baseline speed after each magnet sequence before beginning subsequent magnet sequences; we stopped recording from motors which did not resurrect fully (on average, motors were run 3.8 $\pm$ 2.2 times at each speed before failure).

\section*{Unit unbinding varies with speed near zero torque; binding does not}

We use a simple Hill-Langmuir model for stator assembly \cite{nord2017catch}. We consider the rotor to be surrounded by $N_{\text{max}}$ independent, noninteracting binding sites, each of which can house a single stator unit. A freely diffusing unit binds to an empty binding site with rate $k_{\text{on}}$, and a bound stator unit unbinds with rate $k_{\text{off}}$. We note quickly here that there is no convincing evidence that fixed binding sites exist along the rotor's circumference; a more generalised model could be considered in place of the chosen Hill-Langmuir adsorption. In this model, the random sequential absorption (RSA) model, the periphery of the rotor has no discrete, separated binding spots. This can lead to suboptimal packings when many stators are present -- think, for instance, of a kerb without designated parking spots. At high loads (and high stator unit occupancies), a new stator unit can be blocked from engaging the motor if it is not allotted enough contiguous space on the ring (e.g., due to suboptimal packing of already docked units). As unit occupancy near the zero-torque limit is low, the probability of such a situation is vanishing in this regime; further, previous simulations suggest that differences in relaxation dynamics are negligible between these two models \cite{nord2017catch}. To this end, we focus primarily on the Hill-Langmuir model for ease of analysis.

Under this model, the stator unit occupancy $N(t)$ follows
\begin{equation}
\frac{dN}{dt} = k_{\text{on}}(N_{\text{max}} - N) - k_{\text{off}}N. 
\end{equation}
Motivated by observations that the pool of membrane-bound stator unit `spares' is large \cite{leake2006stoichiometry} ($\sim$200 vs our $N_{\text{max}}$=14 \cite{nord2017catch}), we consider their concentration to be unaffected by motor kinetics, and therefore $k_{\text{on}}$ to be independent of $N$. 

At steady state, $\frac{dN}{dt} = 0$, and we solve for the steady state stator unit occupancy $N_{\text{ss}}$ as
\begin{equation}
N_{\text{ss}} = \frac{N_{\text{max}}}{1+K_D}.
\end{equation}
Here, $K_D = k_{\text{off}}/k_{\text{on}}$ is the disassociation constant. Given evidence that $N_{\text{ss}}$ decreases with decreasing external viscous load, $K_D$ accordingly increases \cite{nord2017catch}. A load-dependent $K_D$ can arise from a load-dependent $k_{\text{on}}$, a load-dependent $k_{\text{off}}$, or both. 

Looking to the solution of (2), together with (3), leads to an expression for the time course of stator occupancy:
\begin{equation}
N(t) = N_{\text{ss}} + (N_0 - N_{\text{ss}})e^{-(k_{\text{on}}+k_{\text{off}})t},
\end{equation}
where $N_0$ is the observed stator unit occupancy before manipulation. In fitting Equation 4 to our data in Figure 4, we make the assumption that $k_\text{off}$, as well as $k_\text{on}$, is not a function of stator unit number. This is consistent with the assumption that $k_\text{off}$ is a function only of the mechanical strain in each unit, proportional to the torque generated per unit, which is not expected to change in our experiments (see Figure 3b). While there have been conflicting results on this in the low load regime \cite{nord2017speed,wang2017limiting}, we are encouraged in our selection of a simple model as our data (apart from at 51 Hz, which does not result in a notable change in $N$) is well fit by a single exponential (Figure 4).

We note that this assumption holds across the torque-speed curves on the right of Figure 3b, but fails for the set of curves on the left when the torque per stator changes drastically at low stator number and high speed --- for instance, from two stator units to one. More detailed exploration in this regime, likely at finer time resolution than is sampled here, will be needed to concretely distinguish between these two scenarios.

We define $t_c = 1/(k_\text{on} + k_\text{off})$ as the fitted decay time constant (Figure 4). From here, binding $k_{\text{on}}$ and unbinding $k_{\text{off}}$ rates are given by:
\begin{equation}
k_{\text{on}} = \frac{N_{\text{ss}}}{t_cN_{\text{max}}},
\end{equation}
\begin{equation}
k_{\text{off}} = \frac{N_{\text{max}} - N_{\text{ss}}}{t_cN_{\text{max}}}.
\end{equation}
In the ranges of load considered by the authors, Nord et al. found that $k_{\text{off}}$ increases with decreasing load, while the binding rate $k_{\text{on}}$ stays largely constant. Stator number did not change appreciably in the 51 Hz condition (Figure 4), and so this time course could not be fit well to Equation 4. Therefore, in this condition, we chose $k_{\text{on}}$ to be the average of the binding rates at higher speeds (101 Hz, 201 Hz, 301 Hz). This was done because our calculated binding rate seemed to remain constant across our measurements. Then, knowing $N_\text{ss}$, we used Equations 5 and 6 to solve for $k_\text{off}$. These calculations are as performed in \cite{nord2017catch}.

We explore this trend in the low load regime and find, consistent with Nord et al., that the unbinding rate increases as motor speed increases (and accordingly, load decreases; see Figure 5c). The binding rate $k_{\text{on}}$ does not change significantly with load (Figure 5b). This is consistent with that of a catch-bond: as the external load decreases, so decreases the stability and the lifetime of the bond.

\section*{IMF likely works via motor torque to affect stator dynamics}

Though much of the recent work on remodelling in the motor has focused on external load, early indications of the dynamic nature of the BFM's stator came courtesy of investigations into IMF effects \cite{khan1980steady,Meister1987,fukuoka2009sodium,Tipping2013}. Here, we revisit the role of IMF (here, PMF, as we are dealing with wild-type H$^+$-driven \textit{E. coli} motors) in stator unit kinetics. We look at remodelling kinetics at the same magnet-driven motor speeds in a low-concentration butanol solution to lower the PMF as in \cite{krasnopeeva2018single}; in this way, we consider the question of whether IMF works independently of load to affect motor assembly, or if motor energetics act indirectly through mechanics in the remodelling process.

To lower the IMF is to take available energy away from the motor, and a natural effect of de-energisation is a lowering of motor speed and torque \cite{lo2013mechanism}. We observe this directly: the single-stator speed in butanol solution is significantly lower than that in standard motility buffer (1.9 $\pm$ 0.9 Hz  vs 3.7 $\pm$ 1.0 Hz in plain motility buffer). For a given rotational speed imposed on the motor via the electromagnets, we expect the motor's torque in a butanol solution to be lower than that in motility buffer; indeed, we observe higher unbinding rates $k_{\text{off}}$ in butanol than in plain motility buffer for all speeds; these differences are more dramatic until the zero-torque speed is reached (likely between 200 and 300 Hz; Figure 5a,c). Changes in the binding rate $k_{\text{on}}$ are not significant (Figure~5b). 

Our results, showing that motors rotated at the same speed at low IMF exhibit higher $k_{\text{off}}$ rates (Figure 5c), are in line with the pattern expected for IMF working via motor torque to affect stator dynamics. Further characterisation of the effect of butanol on motor torque is required (including, for instance, full torque-speed curves in butanol) to fully understand the effect of IMF on stator dynamics; we leave this for future work.

\section*{Discussion}

Developing methods to manipulate the external torque on the flagellar motor is necessary to explore several aspects of the motor's functionality fully. The first enquiries into the dynamics of the BFM were performed using tethered cell and bead assays, both of which do not allow, on their own, for a wide range of loads to be studied on a single motor \cite{silverman1974flagellar,ryu2000torque}. The need for a method to manipulate the external load on individual motors became stronger with the discovery that the stoichiometry of the motor's stator is load-dependent: at higher loads, the motor contains a higher number of engaged stator units (up to at least 11), but can maintain only one or two at low loads \cite{Lele2013,Tipping2013}.

We have presented an experimental framework capable of probing dynamic load-dependent remodelling in the bacterial flagellar motor across the entire range of loads the motor can experience, in particular at low load, a region of the torque-speed curve which has thus far proven difficult to reach \cite{lo2013mechanism,nord2017speed}. Our work was performed concurrently with an investigation of this regime using electrorotation \cite{wadhwa2019torque}. The results of this work and those of Wadhwa et al. are complementary: we both observe a load-dependent $k_\text{off}$, which is consistent with the catch bond hypothesis \cite{nord2017catch}. Quantitatively, their estimated `knee' and zero-torque speeds are both higher than we observe (e.g., we report stator unit drop-off at 101 Hz, suggesting a `knee' somewhere between 51-101 Hz); the torque-speed relationship is dependent on several complex factors \cite{chen2000torque,Berg2003,nord2017speed}, and these differences can arise from several origins, including bacterial strain choice. Electrorotation and speed manipulation using an external magnetic field are complementary methods to explore the BFM's dynamics across its full range of operating conditions: electrorotation works by directly varying the torque on the motor \cite{berry1995mechanical}, while magnetic rotation works by controlling the motor speed. In conjunction, these methods have the ability to allow us to investigate, from both sides, the BFM's torque-speed relationship. 

The development of these methods allow us to dynamically manipulate single motors, and therefore to remove some of the variability that arises from making population measurements. Electrorotation has been shown to be quite harsh on cells, due to heating and large lateral forces on the motor \cite{berg1993torque,berry1995mechanical}. Though heating likely can be mitigated using a cooling system \cite{wadhwa2019torque}, magnetic rotation setups are more easily constructed and replicated. The future of such systems is promising: as superparamagetic microbeads with higher magnetisability are developed \cite{van2015biological}, magnetic rotation can be used to push the motor far past its zero-torque speed, as well as in the reverse CW direction, while avoiding the pitfalls that have previously been demonstrated with electrorotation \cite{berg1993torque,berry1996torque,berry1997absence}.

Previous studies of remodelling have shown that mechanosensitivity in the flagellar motor is consistent with that of a catch bond, a bond whose lifetime increases with external load \cite{nord2017catch}. In the BFM, this manifests as a load-dependent unbinding and load-independent binding rate. We observe similar trends in the zero-torque regime (Figure 5); our investigation into the effects of IMF provide further initial support for a load-dependent catch bond mechanism. The rates we calculate (both for unit binding and unbinding) are significantly higher than those calculated in \cite{nord2017catch}. It is likely that at the small timescales considered here, disengaged stators have not yet diffused away from the rotor and may rebind often and transiently \cite{shi2019hidden}. The effects of this process are likely also more easily observed in the low load regime, when there is more `open space' on the rotor's periphery.

A natural extension of our work is to relate flagellar motor speed to bacterial swimming speed. While studies at higher load lend themselves naturally to understanding surface sensing \cite{Lele2013}, understanding motor adaptation at low loads may give us further insight into the torque regime in which free swimming bacteria reside. 

\section*{Author Contributions}  J.A.N., A.L.N., and R.M.B. formulated the hypotheses and designed the experiments; J.A.N. carried out the experiments, analysed data, and wrote the manuscript; J.A.N., A.L.N., and R.M.B. edited the manuscript. 

\section*{Acknowledgments} We thank C.J. Lo and T.S. Lin for help with initial experiments, S.E. Tusk for assistance in microscope assembly and technical troubleshooting, and all the above for useful discussion and comments. 

\section*{Data Accessibility} All data and analysis code is available at http://users.ox.ac.uk/~phys1213 and/or on request from the authors. 

\section*{Funding} J.A.N. was supported by a Fellowship from All Souls College at the University of Oxford, a James S. McDonnell Foundation Fellowship for Studying Complex Systems, and a Fellowship from the Rockefeller University. 
A.L.N. acknowledges funding from the European Research Council under the European Union's Seventh Framework Programme (FP/2007–2013)/ERC Grant Agreement No. 306475. CBS is a member of the France-BioImaging (FBI)
and the French Infrastructure for Integrated Structural Biology (FRISBI), 2 national infrastructures supported by the French National Research Agency (ANR-10-INBS-04-01 and ANR-10-INBS-05, respectively). R.M.B was supported by the Biotechnology and Biological Sciences Research Council grant BB/N006070/1.

\bibliography{ref}
\bibliographystyle{plain}

\newpage

\begin{figure}[t]
\begin{center}
\includegraphics[width=\textwidth]{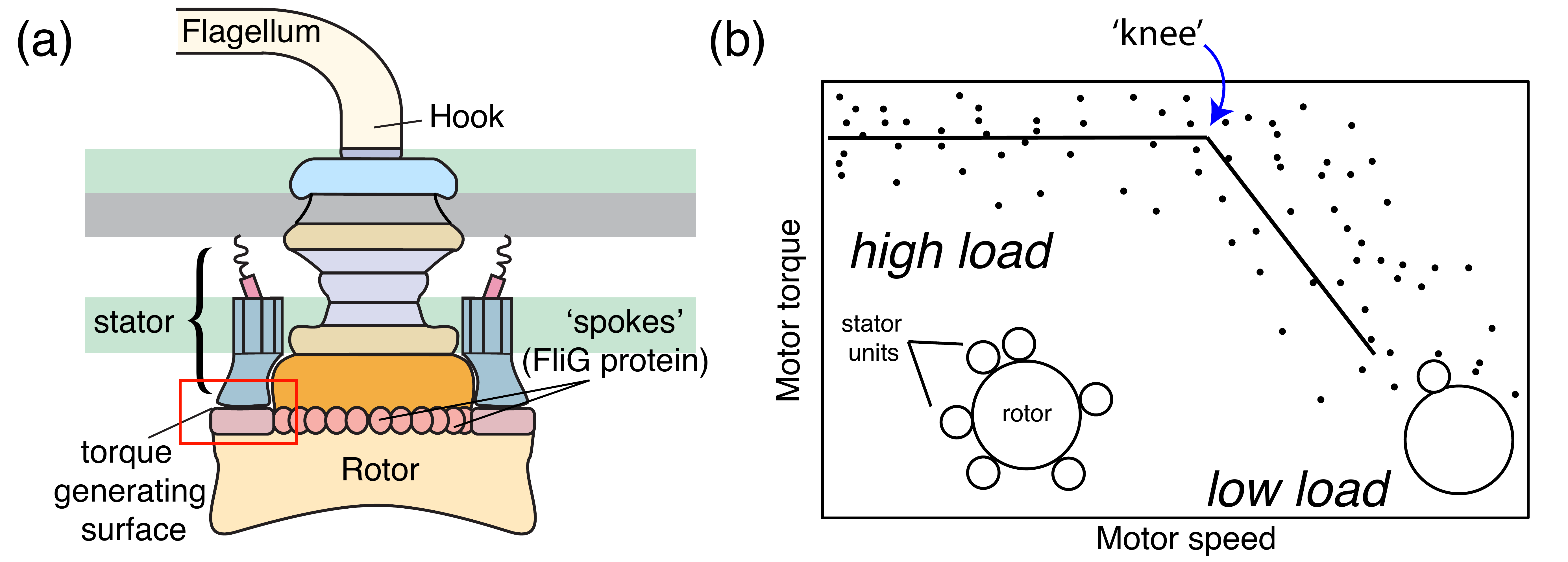}
\end{center}
\caption{BFM structure and dynamics. \textbf{(a)} The flagellar motor's rotor consists of a series of large co-axial rings that attach to a flagellar filament via a flexible hook. An active motor can have up to at least 11 torque-generating stator complexes. Stators interact with proteins (FliG) along the rotor's edge to drive motor rotation. \textbf{(b)} Experiments in recent years have established that the number of torque-generating units varies with external load on the motor (among other possible factors, including IMF). Points in the high-load regime correspond to motors near full occupancy and points at low loads to motors with only one or two. Data shown from \cite{chen2000torque}. Solid lines are included to guide the eye.}
\end{figure}

\begin{figure}[t]
\begin{center}
\includegraphics[width=\textwidth]{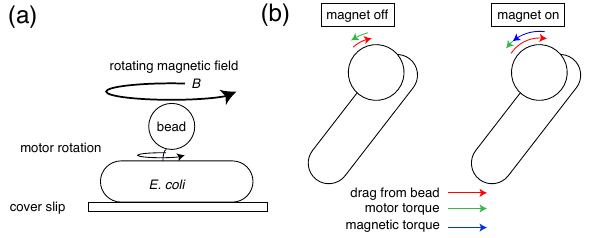}
\end{center}
\caption{Manipulating motor speed using a rotating magnetic field. \textbf{(a)} Experimental setup: Bacteria are immobilised on a coverslip, and a superparamagnetic bead is attached to the motor's hook; (x,y) positions of the bead are tracked and used to obtain speed vs time traces. Current flowing through a three-pole electromagnet generates a rotating magnetic field $B$ to manipulate the speed of the motor. \textbf{(b)} When the magnet is turned off (left), motor torque (green) must balance viscous drag from the bead (red), resulting in a low rotation speed. When the magnetic field is on and rotating (right), the magnetic torque (blue) takes on most of the drag from the bead, allowing for faster rotation with minimal contribution from motor torque. View is from the top of a CCW rotating motor.}
\end{figure}

\begin{figure}[t]
\begin{center}
\includegraphics[width=\textwidth]{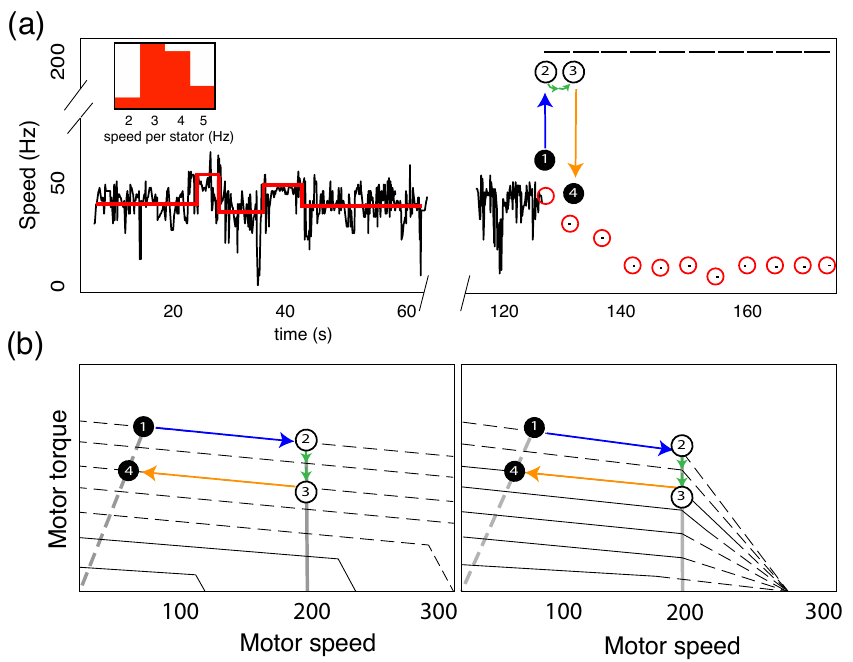}
\end{center}
\caption{\textbf{(a)} An example experimental trace. Motors are allowed to freely rotate for at least 1 minute, during which time a step detection algorithm is used to calculate single stator speeds and determine stator unit stoichiometry (red lines). 
An inlay shows a histogram of single stator speeds for the plain motility buffer condition. Motors are then rotated at a fixed speed (201 Hz in this example) for 5 second intervals with 0.2s `off' periods during which the motor speed is recorded. `Magnet on' periods are shown as lines at 201 Hz; red circles depict an example time course of motor remodelling;each red circle contains a single speed calculated within one FFT window. The results of the first `magnet on' period are shown by points 1,2,3,4. At point 1, the motor has 7 stator units. When the magnetic field rotates the motor at 201 Hz (blue arrow to point 2), the motor loses 2 stator units (green arrows) between points 2 and 3 (open circles, unobserved). This loss is observed as a decrease in speed during the first `magnet off' period at point 4.  \textbf{(b)} Two possible torque-speed curves for a motor with 7 stator units. The left depicts results by \cite{nord2017speed}, which showed that the zero-torque speed of the BFM increased with the number of engaged stator units; the right depicts results by \cite{ryu2000torque}, which suggested that the limiting speed is independent of stator number. Solid lines are based on data; dashed lines are extrapolated as in \cite{nord2017speed} and \cite{ryu2000torque}. Dashed grey lines are load lines for the passive beads; solid grey lines show constant speed across stator number during magnet-assisted rotation. Recent experiments by \cite{sato2019evaluation} have supported the results of \cite{nord2017speed} (left). Points 1,2,3,4 depict the same cycle of the magnet from the experimental trace shown in \textbf{(a)}. Our assumption that unit binding and unbinding rates are not a function of stator number holds across the torque-speed curves on the right. Considering the left set of curves, this assumption fails when the torque per stator changes drastically at low stator number and high speed --- for instance, from two stator units to one (see solid grey line). Further exploration in this regime, likely at finer time resolution than is sampled in this work, will be needed to distinguish between these two scenarios.} 
\end{figure}

\begin{figure}[t]
\begin{center}
\includegraphics[width=\textwidth]{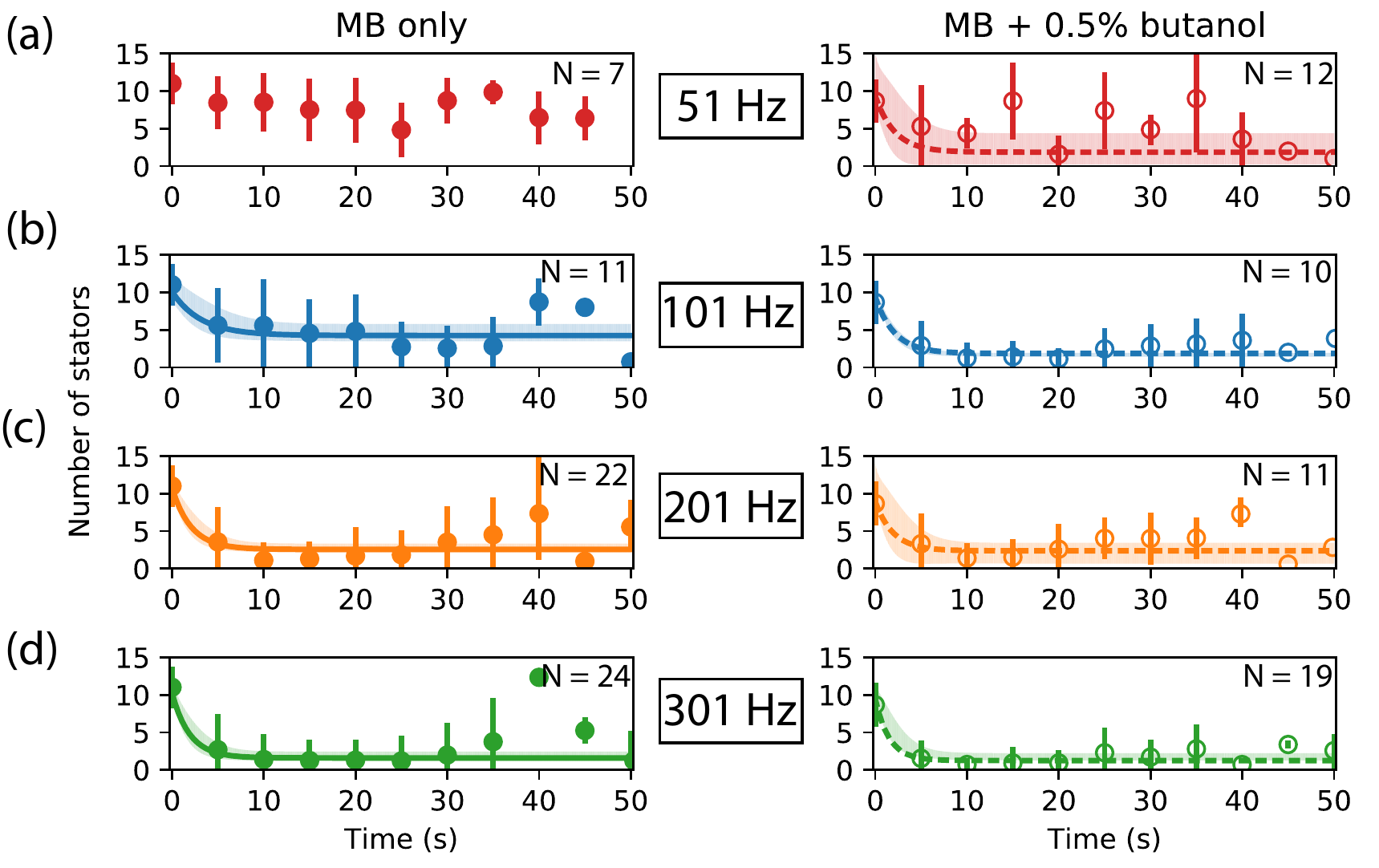}
\end{center}
\caption{Motor remodelling following a sudden change in load in plain motility buffer (left) and at low PMF (right). Low PMF environment is attained by mixing 0.5\% butanol, shown to act as an ionophore \cite{krasnopeeva2018single}, into motility buffer (MB). Time traces show stator unit stoichiometry after magnets are turned on at \textbf{(a)} 51 Hz (red); \textbf{(b)} 101 Hz (blue); \textbf{(c)} 201 Hz (orange) and \textbf{(d)} 301 Hz (green). Circles and error bars show mean $\pm$ standard deviation for experimental data in each condition; filled circles represent data collected in plain MB and open circles data collected in low PMF buffer. Lines (solid for MB condition; dashed for low PMF) and shaded regions depict fits of Equation 4 to data $\pm$ 3 standard deviations. We were not able to fit the traces collected at 51 Hz in plain MB (top left, red), which did not lose stators appreciably during magnetic rotation, to this model.}
\end{figure}

\begin{figure}[t]
\begin{center}
\includegraphics[width=\textwidth]{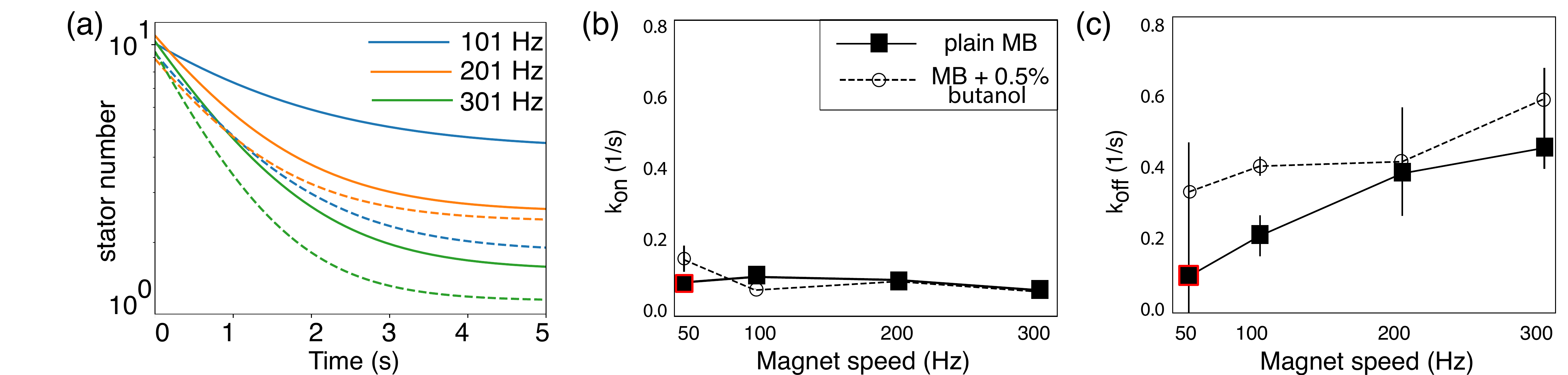}
\end{center}
\caption{Comparison of motor remodelling dynamics at different speeds and IMF levels. \textbf{(a)} A zoomed-in look at the early, most dynamic region from the plots in Figure 4; lines shown are fits of Equation 4 to the data in that figure. As previously, dashed lines represent results in the lowered PMF condition and solid lines in plain MB. The same colour scheme for magnet speed is used throughout this manuscript. We extrapolate the zero-torque speed in the plain MB condition to vary between 201 and 301 Hz, and between 101 and 201 Hz in the butanol condition; dynamics at these speeds are similar, while those at appreciably higher loads (i.e., 101 Hz in MB, solid blue line) and appreciably lower loads (i.e., 301 Hz in butanol + MB, dashed green line) diverge. \textbf{(b)} Binding rates $k_\text{on}$ remain relatively constant in both IMF conditions across all four speed levels while \textbf{(c)} unbinding rates $k_\text{off}$ vary as speed is increased and are on average higher in the lowered IMF condition as compared to in plain MB. Rates $k_\text{on}$ and $k_\text{off}$ were calculated from exponential fits to the data of Figure 4, except for the points outlined in red (see text for details). These independent effects suggest that the effect of both speed and IMF on stator dynamics in the BFM quite likely works via the relationship of these factors to the load on the motor. Our results also show behaviour consistent with that of a catch bond, as did those of a prior investigation in the higher load regime \cite{nord2017catch}.}
\end{figure}

\end{document}